\documentclass[aps,twocolumn,preprintnumbers,floatfix,reprint, nofootinbib]{revtex4-1}
\usepackage{mathrsfs,natbib}
\usepackage{amsmath,amssymb, amsfonts, bbm}
\usepackage{graphics,epsfig,color,subfigure}
\usepackage{hyperref}
\usepackage{pstricks}

\newcommand{\be}{\begin{equation}}
\newcommand{\ee}{\end{equation}}

\newcommand{\dslash}{D\!\!\!\!\slash \,\,}
\newcommand{\Pf}{\mathrm{Pf}}
\def\bea{\begin{align}}
\def\ena{\end{align}}

\def\Tr{\textrm{Tr}}
\def\beqa{\begin{eqnarray}}
\def\enqa{\end{eqnarray}}

\def \Pf {\textrm{Pf}}

\def\slashchar#1{\ensuremath{                               %
   \setbox0=\hbox{${}#1{}$}       
   \dimen0=\wd0                                 
   \setbox1=\hbox{/} \dimen1=\wd1               
   \ifdim\dimen0>\dimen1                        
      \rlap{\hbox to \dimen0{\hfil/\hfil}}      
      {}#1{}                                    
   \else                                        
      \rlap{\hbox to \dimen1{\hfil${}#1{}$\hfil}}   
      /                                         
   \fi}}

\begin{document}
\title{Large $N_{c}$ Equivalence and Baryons}

\author{Mike Blake}
\email{M.A.Blake@damtp.cam.ac.uk} 
\author{Aleksey Cherman}
\email{A.Cherman@damtp.cam.ac.uk} 

\affiliation{Department of Applied Mathematics and Theoretical Physics, University of Cambridge, Cambridge, CB3 0WA, United Kingdom}
\preprint{DAMTP-2012-28}

\begin{abstract}
In the large $N_{c}$ limit, gauge theories with different gauge groups and matter content sometimes turn out to be `large $N_{c}$ equivalent', in the sense of having a set of coincident correlation functions.   Large $N_{c}$ equivalence has mainly been explored in the glueball and meson sectors.   However, a recent proposal to dodge the fermion sign problem of  QCD with a quark-number chemical potential using large $N_{c}$ equivalence motivates investigating the applicability of large $N_{c}$ equivalence to correlation functions involving baryon operators.   Here we present evidence that large $N_{c}$ equivalence extends to the baryon sector, under the same type of symmetry-realization assumptions as in the meson sector, by adapting the classic Witten analysis of large $N_{c}$ baryons. 
\end{abstract}

\maketitle
\newpage

\section{Introduction}

Large $N_{c}$ limits of gauge theories\cite{Hooft:1973jz,Witten:1979kh} have many remarkable and beautiful properties. A particularly striking property of some large $N_{c}$ gauge theories is the existence of  large $N_{c}$ orbifold equivalence\cite{Kachru:1998ys,Bershadsky:1998cb,Kovtun:2003hr,Kovtun:2004bz}, which implies  that gauge theories with different matter content and gauge groups can have a subset of coincident correlation functions\footnote{In the pure gauge case such relationships were discovered very early on\cite{Cicuta:1982fu,*Lovelace:1982hz}.}.  Such equivalences can be useful if one of the equivalent theories is more tractable than another, as in the well-known case of the `orientifold equivalence' discussed in \cite{Armoni:2003gp,Armoni:2004uu}.   To date, explorations of large $N_{c}$ equivalences have focused on correlation functions involving glueball and meson operators, which couple to states that remain light in the large $N_{c}$ limit\footnote{For an exception, see e.g. \cite{Armoni:2010ny}.}.  In this paper we discuss the extension of large $N_{c}$ equivalence to baryons, in the context of a particular topical example. 


The motivation for exploring the applicability of large $N_{c}$ equivalences to baryons is a recent proposal that a large $N_{c}$ equivalence might be useful for studying QCD at finite density\cite{Cherman:2010jj}, by providing a way to dodge the so-called sign problem of lattice QCD.    The sign problem refers to the issue that when a chemical potential for quark number $\mu \bar{\psi} \gamma^{0} \psi$ is turned on, the fermion determinant of $SU(N_{c})$ QCD becomes complex.  This makes Monte Carlo evaluation of the path integral of QCD impossible, because Monte Carlo methods rely on using importance sampling with $Z^{-1}\det(\dslash + m_{q} + \mu \gamma^{0})\exp(-S_{YM})$ as a probability distribution, where $S_{YM}$ is the glue action, and $Z$ is the partition function, and a probability distribution must be positive.   The sign problem has been notoriously intractable for $N_{c}=3$ QCD, prompting attempts to find other, related, theories where one may be able to make some progress. With this motivation, \cite{Cherman:2010jj} proposed an approach to dodge the sign problem of QCD by studying an orbifold-equivalent theory, with the QCD gauge group $SU(N_{c})$ replaced by the gauge group $SO(2N_{c})$, which has been further explored in \cite{Hanada:2011ju,Cherman:2011mh,Hidaka:2011jj,Hanada:2012nj}.  In these works, the focus was on correlation functions involving meson and glueball operators.  However, especially given the motivations of this proposal in the physics of baryons at finite density, it is natural to ask whether this large $N_{c}$ equivalence applies to correlation functions of baryons as well. 

To see why the baryon case is subtle, we first briefly summarize how the orbifold equivalence works for meson correlation functions.   The equivalence is supposed to hold in the the `t Hooft large $N_{c}$ limit, where one takes $N_{c} \to \infty$ with the `t Hooft coupling $\lambda \equiv N_{c} g_{YM}^{2}$ and $N_{f}$ held fixed.  If $m(x)$ is a meson operator at position $x$ to which equivalence applies (a `common-sector' operator), then at large $N_{c}$ we expect an equivalence of (connected) correlators of $m(x)$ between $SU(N_{c})$ gauge theory with `t Hooft coupling $\lambda$ and $SO(2N_{c})$ gauge theory with 't Hooft coupling $2\lambda$,

\begin{align}
 \langle m (x_1)...m(x_n) \rangle_{SO(2N_{c}), 2\lambda} =  2 \langle m(x_1)...m(x_n) \rangle_{SU(N_{c}), \lambda} 
 \label{eq:equivalence}
 \end{align}
 This relation implies (for instance) that the masses of the lightest mesons $m$ coupling to $m(x)$ will agree in the two theories, since the masses can be read off from the large-time $t$ behavior of the two-point correlation function $\sim  e^{-m t}$ after Wick rotation to Euclidean space. However the orbifold equivalence relates gauge theories with different numbers of colors, and so baryons in the two theories will be constructed from a different number of quarks.  Since baryon masses depend non-trivially on the number of colors in the large $N_{c}$ limit, baryon masses cannot coincide in the two theories.  Hence we should not expect a relation of the form ~\eqref{eq:equivalence} to apply in the baryonic sector.  It is therefore clear that a direct use of this orbifold equivalence, which relates theories with different number of colors, will not be appropriate for extending the large $N_{c}$ equivalence to baryons. 

Fortunately, this observation does not necessarily imply that large $N_{c}$ equivalence gives no relationship between baryon correlation functions, because there is a natural way to deal with the problem we described above. In the large $N_{c}$ limit the dependence of meson correlation functions on $N_{c}$ is just a simple scaling. Therefore the above orbifold equivalence can easily be reinterpreted as defining an equivalence of neutral mesons between $SU(N_{c})$ and $SO(N_{c})$ gauge theory. This provides a more natural starting point for discussing the extension of large $N_{c}$ equivalence to baryons. Although we can no longer directly use the technique of orbifold projections, the equivalence of neutral sector operators can be seen simply in perturbation theory. Furthermore, such an approach is equally valid for identifying large $N_{c}$ equivalences between baryons. Hence our approach in most of this paper is to give up the elegance of working directly with the formal language of orbifold projections and instead to explore the large $N_{c}$ equivalence between $SO(N_{c})$ and $SU(N_{c})$ gauge theories directly.

The organization of this paper is as follows. First, we briefly review the large $N_{c}$ orbifold equivalence between $SO(2N_{c})$ and $SU(N_{c})$ gauge theories in Sections~\ref{sec:OrbifoldProjections} and explain the difficulties in using it to discuss baryons in more detail. In Section~\ref{sec:MesonEquivalence} we reformulate the proof of large $N_{c}$ equivalence for meson correlation functions in theories with the same number of colors directly in terms of Feynman diagrams.  We then show how this argument allows us to generalise large $N_{c}$ equivalence, at least in perturbation theory, to baryons in Sections~\ref{sec:BaryonEquivalence} and~\ref{sec:BaryonInteractions}. In particular we show that the masses of 'common' baryons should agree at leading order in $N_{c}$. To get some insight into the non-perturbative conditions for the validity of the large $N_{c}$ equivalence, in Section~\ref{sec:2D} we discuss 2D QCD with gauge groups $SO(N_{c})$ and $SU(N_{c})$, where one can go beyond perturbation theory and work directly with the full path integral.   Finally, in Section~\ref{sec:Conclusions} we summarize our findings and sketch some possible directions for future work. 

\section{Orbifold projections}
\label{sec:OrbifoldProjections}
In this section we briefly review the orbifold projection that connects $SO(2N_{c})$ gauge theory and $SU(N_{c})$ gauge theory. Orbifold projections give algorithms for constructing gauge theories which may be large $N_{c}$ equivalent.  The idea is to start with a given `parent' theory, and apply a projection based on some symmetry of the action. One then constructs a `daughter' theory by discarding all the degrees of freedom that transform non-trivially under the chosen symmetry. Then one can show that under certain conditions\cite{Kovtun:2003hr,Kovtun:2004bz}, certain correlation functions coincide between the parent and daughter theories at leading order in $N_{c}$ to all orders in perturbation theory \cite{Bershadsky:1998cb}. In particular these `common sector' operators include the mesons operators (i.e. gauge invariant quark bilinears) in the parent theory that are invariant under the chosen symmetry and their orbifold projections in the daughter. The necessary and sufficient conditions for these equivalences to hold non-perturbatively are not fully understood outside the context of gauge theories with fermions in two-index representations\cite{Kovtun:2003hr,Kovtun:2004bz}. However, it is understood that in order for an orbifold equivalence to hold the critical necessary condition is that the symmetry used in the projection must not be spontaneously broken. 

It was recently pointed out that $SU(N_{c})$ and $SO(2N_{c})$ gauge theories with fundamental Dirac fermions are related by an orbifold projection. The orbifold projection relating an $SO(2N_{c})$ with $N_{f}$ flavors of Dirac fermions to an $SU(N_{c})$ gauge theory with $N_{f}$ fermions, which is QCD, is based on the $\mathbb{Z}_{2}$ symmetry defined by taking an element $J\in SO(2N_{c})$ given by $J = i\sigma_{2} \otimes 1_{N_{c}}$, with $1_{N}$ is an $N \times N$ identity matrix, and  $\omega = e^{i \pi/2} \in U(1)_Q$. Here $U(1)_Q$ is the quark number symmetry that acts on the quark fields as $\Psi \rightarrow \omega \Psi$.  $J$ generates a $\mathbb{Z}_{4}$ subgroup of $SO(2N_{c})$ and $\omega$ generates $\mathbb{Z}_{4} \in U(1)_{Q}$, but since the action of $J$ and $\omega$ on the $SO$ gauge field $A_{\mu}$ and the fermion field $\Psi$ is
\be
\label{eq:OrbifoldProjectionStandard}
A_{\mu} \rightarrow J A_{\mu} J^{T}, \;\; \Psi \rightarrow \omega J \Psi,
\ee
the combined action of $J,\omega$ is a $\mathbb{Z}_{2}$ symmetry.  Some algebra shows that a projection of the $SO(2N_{c})$ theory based on this $\mathbb{Z}_{2}$ gives $SU(N_{c})$ QCD as the daughter theory, with an equivalence expected in the 't Hooft large $N_{c}$ limit so long as the $\mathbb{Z}_{2}$ projection symmetry is not spontaneously broken\cite{Cherman:2011mh,Hanada:2012nj}.   The key feature of this particular orbifold projection is that it can be applied in the presence of a quark number chemical potential in the $SO(N_{c})$ theory, and the result is an $SU(N_{c})$ theory with a quark-number chemical potential, and the latter theory is of obvious phenomenological interest.


The key symmetry used in the orbifold projection is based on the quark number charge $Z_{4} \in U(1)_Q$. In $SO(N_{c})$ gauge theory we have two types of color singlet quark bilinears: `mesons', of the form $\bar{\Psi} \Gamma \Psi$; and `b-mesons' $\Psi^{T} C \Gamma \Psi$ with $\Gamma = 1,\gamma_{\mu},\ldots$. The b-mesons are charged under quark number and so are not in the common sector. Therefore the orbifold equivalence is a correspondence between the mesons in $SO(2N_{c})$ and $SU(N_{c})$ gauge theories. This equivalence is expected to hold in the 't Hooft large $N_{c}$ limit  so long as the $\mathbb{Z}_{4}\in U(1)_{Q}$ remains unbroken. In this paper we will mainly discuss the physics with $\mu=0$, where the Vafa-Witten\cite{Vafa:1983tf} theorem implies that $U(1)_{Q}$ cannot break spontaneously, and hence the equivalence should hold.  For general $\mu$ the story is more subtle and is discussed in \cite{Cherman:2010jj,Cherman:2011mh}.

In  $SU(N_{c})$, `quark number' and `baryon number' are often used interchangeably, since the only states charged under quark number are the baryons, the interpolating operators for which contain $N_{c}$ quark fields contracted with a color-epsilon tensor. Hence at non-zero chemical potential, the daughter theory has a finite density of baryons. It is therefore natural to ask if these baryons are related to operators in the $SO(2N_{c})$ theory. In the $SO(2N_{c})$ gauge theories, using  `quark number' and `baryon number' interchangeably would be unfortunate, since there are the gauge-invariant `b-meson' states composed of two valence quarks which carry quark number $\pm 2$. Similarly in an $SO(2N_c)$ theory one can obtain gauge-invariant states by contracting both quark and antiquark fields with the same epsilon tensor, since quarks and antiquarks transform the same way under color.  Hence the quark number charge of baryons --- that is, states whose interpolating operators involve epsilon tensors --- can be less than $|2N_{c}|$.  

There are therefore more baryon states in the $SO(2N_c)$ gauge theory than in the $SU(N_c)$ theory. By analogy with the meson case, we expect a subset of these baryons to correspond to those in the $SU(N_c)$ theory. A natural candidate for such a `common sector' baryon might be the baryon operator, $B$,  composed entirely of quarks

\begin{align}
B = \epsilon_{i_{1} i_{2} \cdots i_{2N_{c}}} \Psi^{i_{1}}\Psi^{i_{2}}\cdots \Psi^{i_{2N_{c}}}
\end{align}
where $i_{j}$ are color indices, and flavor and spin indices are suppressed. In particular, for even $N_{c}$ this operator is neutral under $\mathbb{Z}_{4}\in U(1)_{Q}$.


However, if one actually applies the projection of Eq.~\eqref{eq:OrbifoldProjectionStandard} in the standard way, $B$ is annihilated. The reason is that this orbifold projection discards half of the color degrees of freedom, leaving only $N_{c}$ colors for the projected quark fields.  The color sum involving the $2N_{c}$-index $\epsilon$ tensor in Eq.~\eqref{eq:OrbifoldProjectionStandard} then gives zero, suggesting that baryons are \emph{never} in the common sector. 

Standard large $N_{c}$ arguments suggest that this is a rather counter-intuitive conclusion. At large $N_{c}$ Witten's classic analysis showed that baryons can be interpreted as solitons of meson fields. Since the properties of the common-sector (neutral) mesons coincide in the two theories, one would expect to be able to construct a soliton of neutral mesons in $SO$ gauge theory which should be identifiable with the $SU$ baryon. Such considerations suggest that the direct application of the orbifold projection recipe to operators which contain $\epsilon$ tensors, such as $B$, may be misleading.  

As we discussed in the introduction, it appears that a major part of the difficulties with discussing baryons using this orbifold projection can be traced to the fact it relates theories with different numbers of colors\footnote{It is possible to obtain $SO(N_{c})$ gauge theory with Majorana fermions via an orbifold projection of $SU(N_{c})$ with Dirac fermions using charge conjugation, and it may be in such a context the language of orbifold equivalence may be more directly useful for discussing baryons. However, such a projection does not make sense at finite chemical potential since the chemical potential breaks the charge conjugation symmetry, and so understanding whether baryons survive such projections would not tell us directly whether they survive the phenomenologically-motivated projection we focus on in this paper. We also note that it is possible to discuss the large $N_{c}$ equivalences between $SU(N_{c})$ and $SO(N_{c})$ with Dirac fermions theories using the orbifold language by viewing the $SO(N_{c})$ theory as a projection of the $SO(2N_{c})$ by $J = 1 \otimes \sigma_{3} \in SO(2N_{c})$.  Then the $SU(N_{c})$-$SO(N_{c})$ equivalence is a \emph{daughter-daughter} orbifold equivalence.  However, we have not found a  way to use this point of view to clarify the issues with baryons, and hence will not discuss it further. }. This was not a problem when discussing mesons and glueballs because those correlation functions have a trivial dependence on $N_{c}$. Baryons, on the other hand, are made from $N_{c}$ quarks and so we can only expect to construct direct equivalences between theories with the same number of colors. It is therefore not surprising that a direct application of the orbifold equivalence is unsuccessful.

\section{Large $N_{c}$ equivalence in perturbation theory}

Given the difficulties with a direct application of the standard orbifold prescription to baryon-sector operators, we shall explore large $N_{c}$ equivalence between $SU(N_{c})$ and $SO(N_{c})$ theories by explicitly comparing Feynman diagrams in the two theories. Whilst perturbative proofs of the meson equivalence have been given before\cite{Bershadsky:1998cb}, we first reproduce these results without reference to the orbifold projection. We then show that this method allows us to successfully generalize the notion of the large $N_{c}$ equivalence to incorporate baryons. 

\subsection{Mesons}
\label{sec:MesonEquivalence}
Since our interest is in applying large $N_{c}$ equivalence to baryons,  we will compare $SO$ and $SU$ gauge theories with the same number of colors. In the large $N_{c}$ limit the difference between $SO(2N_{c})$ and $SO(N_{c})$ is simply a scaling, and so we expect that planar equivalence for meson correlation functions should take the form
\begin{align}
 \langle  m(x_1)...m(x_n) \rangle_{SO(N_{c}), 2\lambda} =  \langle m(x_1)...m(x_n) \rangle_{SU(N_{c}), \lambda} 
 \label{eq:equivalence2}
 \end{align}
where the two gauge theories involved are $SO(N_{c})$ gauge theory with a YM coupling $2g^2$ and $SU(N_{c})$ with coupling $g^2$.  


The only difference in the evaluation of a Feynman diagram in $SO$ and $SU$ gauge theories comes from the traces over color indices. The color structure of the quark propagator and the interaction vertices agree in the two theories. The only difference is in the color structure of the propagators. A free gluon propagator can be written as 
\begin{align}
\langle A_{\mu,j}^{i}(y) A^{k}_{\nu,l}(x) \rangle = D_{\mu \nu}(x-y) C^{i; k}_{j; l}
\end{align}
where $D_{\mu \nu}(x-y)$ is a standard propagator for a massless vector field, and $C$ is the color factor.  

\begin{figure}
\includegraphics{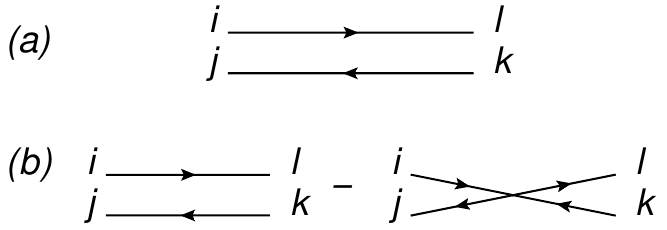}
\caption{\small\emph{The double line notation for (a) $SU(N_{c})$ and (b) $SO(N_{c})$}}
\label{fig:DoubleLinePropagators}
\end{figure}

In $SU(N_{c})$ gauge theory, $C$ takes the form 
\be
\label{eq:SUprop}
C^{i; k}_{j; l} = a (\delta^i_{\;l} \delta^k_{\;j}  - \frac{1}{N_{c}}\delta^i_{\;j} \delta^k_{\;l} )
\ee
where $a = 1/2$.  When $N_{c}$ is large, we can drop the second term, since it is subleading compared to the first term.  Physically, this is simply the statement that at large $N_{c}$ the difference between $U(N_{c})$ gauge theory and $SU(N_{c})$ gauge theory is $1/N_{c}^{2}$ suppressed. The $U(N_{c})$ propagator has the same color structure as a quark-antiquark pair and so can be represented by an arrowed double line as shown in Fig.~\ref{fig:DoubleLinePropagators}. The arrows, which are necessary because the gauge group is complex, represent the direction of color flow. Any Feynman diagram can be redrawn in terms of double lines for gluon propagators and single lines for quark propagators, with a consistent flow of arrows.   

In $SO(N_{c})$ the color structure is:
\be
\label{eq:SOprop}
\tilde{C}^{i; k}_{j; l} =\frac{a}{2}  (\delta^i_{\;l} \delta^k_{\;j}  - \delta^{ik} \delta_{jl})
\ee
with the same $a$. 

We see that the first term in the $SO(N_{c})$ propagator has the same color structure as the propagator in the $SU(N_{c})$ theory, but there is also an additional term which we shall refer to as the `twisted propagator'.   In comparing $SU(N_{c})$ and $SO(N_{c})$ theories, it will be convenient to introduce a somewhat unconventional double line notation for the $SO(N_{c})$ theory. Usually when discussing $SO(N_{c})$ gauge theories, the gluon propagator is represented as a double line, a la 't Hooft, but without any arrows. This reflects the fact that the genus expansion for $SO(N_{c})$ includes non-orientable surfaces, and if one tried to keep track of color flow using arrows in the same way as in the $SU(N_{c})$ case, one would find that there are diagrams for which it is not possible to consistently assign a direction of color flow.  However, it is possible to modify the double line notation such that the $SO(N_{c})$ diagrams also have a modified kind of arrow flow.  Each end of the lines appearing in the double line notation is associated with a color index, with the lines connecting indices which are contracted via Kroenecker delta functions. In our notation for the $SO(N_{c})$ theory,  the arrows on the double lines which indicate whether the external indices associate with each double line are raised or lowered, as shown in Fig.~\ref{fig:DoubleLinePropagators}.   A line emanating from an end associated with an upper index gets an arrow pointing away from the end, while ends associated with lower indices have arrows pointing towards them.   This is a useful notation, because all vertices and propagators actually represent matrix multiplication, which can always be written in terms of raised and lowered indices, and so preserves this `flow' of arrows. One can see that for the usual term in the propagator, this is the same flow as the color flow in $SU(N_{c})$. The twisted propagator couples two upper indices and two lower indices. Diagrammatically, it therefore flips the direction of the arrows and gives the crossed diagram in Fig.~\ref{fig:DoubleLinePropagators}.  As shall be made explicit shortly, this twisted contribution to the propagator is responsible for the introduction of non-orientable diagrams.  

\begin{figure}
\includegraphics{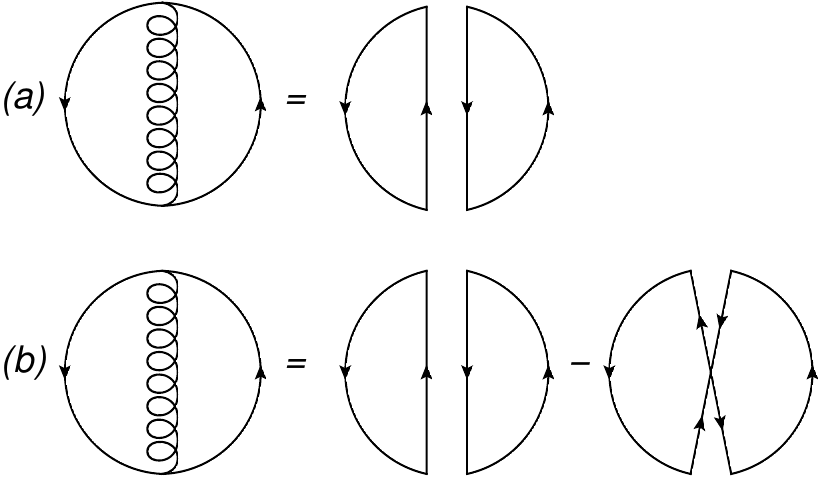}
\caption{\small\emph{The first order diagrams contributing to a common-sector meson correlation function in (a) $SU(N_{c})$ and (b) $SO(N_{c})$. The second diagram in (b) is suppressed by a factor of $N_{c}$}}
\label{fig:OGEMesonCorrelatorDiagram}
\end{figure}

Having established our notation we can now investigate the large $N_{c}$ equivalence between $SO$ and $SU$ theories.  At large $N_c$ we need only consider Feynman diagrams that contain a single quark loop and internal gluons. The meson operator insertions lie on this quark loop. However, from here on we require that these diagrams correspond to \emph{common-sector} operators, which at the Feynman diagram level is simply the demand that a consistent flow of quark number can be assigned to the quark loop. Then the first order diagram for a common-sector meson correlation function is just a quark loop with a single internal gluon, as shown in Fig.~\ref{fig:OGEMesonCorrelatorDiagram}.  In order to compare this process in $SU(N_{c})$ and $SO(N_{c})$ we simply insert the double line propagators onto the diagram such that there is a consistent flow of arrows at the vertices.  We have two diagrams in $SO(N_{c})$ corresponding to the two terms in the propagator. We see that the first term reproduces the same planar diagram as in $SU(N_{c})$. The second term in the propagator corresponds to a non-orientable diagram which contains only one index loop and so is suppressed by a factor of $N_{c}$. The only difference between the two theories in the large $N_{c}$ limit is the factor of two in the normalization of the propagator, which is precisely compensated for by the difference in the coupling constants.  

\begin{figure}
\includegraphics{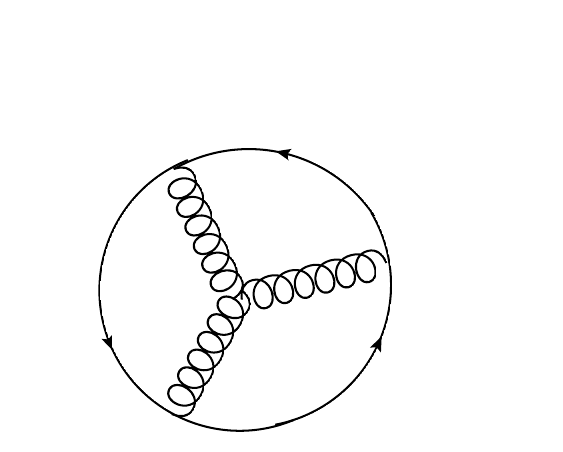}
\caption{\small\emph{The three gluon interactions contains three gluon propagators and four coupling constants. This tells us that simply counting these factors is not enough to understand the orbifold equivalence}}
\label{fig:ThreeMesonVertex}
\end{figure}

As was illustrated in our example the only double line diagrams that survive in the large $N_c$ limit are those `planar' diagrams where the gluons tile the quark loop. We wish to argue that the meson correlators match for all such planar diagrams. Before coming to the general argument, it is instructive to first consider one further example - a quark bubble with an interaction mediated via the three-gluon vertex, illustrated in Fig.~\ref{fig:ThreeMesonVertex}. The diagram contains $4$ vertices, each bringing a factor of the coupling constant and three propagators. Therefore if one were to make a tempting generalization from the preceding example and naively make the assumption that the use of the `twisted propagator' in $SO(N_{c})$ only results in subleading diagrams, one would conclude that the ratio of the amplitude associated to Fig.~\ref{fig:ThreeMesonVertex} in the $SO(N_{c})$ theory to the amplitude in the $SU(N_{c})$ theory is $2^{4/2} \times 2^{-3} = \frac{1}{2}$, suggesting a mismatch. 

\begin{figure}
\includegraphics{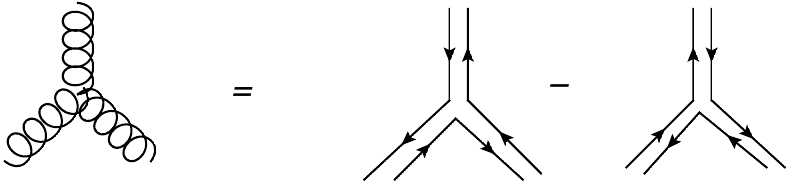}
\caption{\small\emph{The three gluon vertex in the double line notation can come in two orientations.}}
\label{fig:3gluonvertex}
\end{figure}

The resolution of this apparent paradox is that we can in fact use the twisted propagators to construct a second equivalent planar diagram. This is because in both theories, the three-gluon vertex corresponds not to one but to two vertices in the double line notation Fig.~\ref{fig:3gluonvertex}. These two vertices have different orientations - one has the same orientation as the external quark loop, the other, the `twisted vertex', has the opposite orientation, and comes with a relative minus sign. In $SU(N_{c})$ gauge theory we can only construct a planar diagram from the vertex that has the same orientation as the external loop, since otherwise we would not have a consistent flow of arrows. 

\begin{figure}
\includegraphics{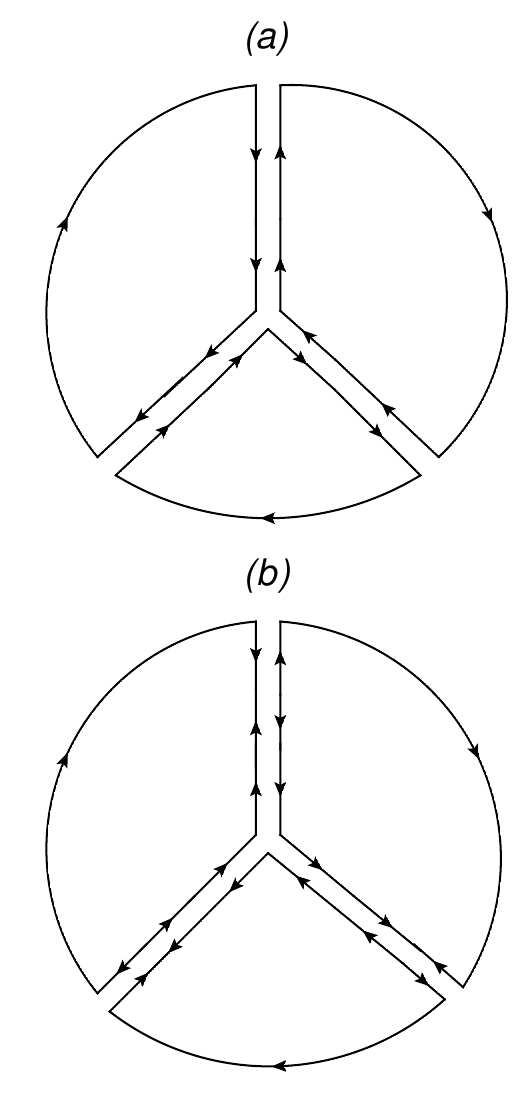}
\caption{\small\emph{In $SU(N_c)$ gauge theory we can only construct planar diagrams from vertices that have the same orientation as the external quark loop, e.g. diagram (a). In $SO(N_c)$ gauge theory we can use to the twisted propagators to also construct diagram (b), where the vertex has the opposite orientation.}}
\label{fig:3GAdditionalSODiagram}
\end{figure}

On the other hand, in the $SO(N_{c})$ gauge theory we can effectively reverse the orientation of a vertex by attaching enough twisted propagators to it. In this way we can construct a second planar diagram which consists of the twisted vertex and three twisted propagators (Fig.~\ref{fig:3GAdditionalSODiagram}). The minus signs in the propagators and the twisted vertex give a factor of $(-1)^{3}(-1)$ and so this diagram is identical to the usual planar one. At large $N_{c}$ the ratio of any interaction in the two theories therefore has three contributions arising from the relative number of planar diagrams, the coupling constants and the propagators. For the three-gluon interaction the factors of $2$ combine so that there is an equivalence between the $SO$ and $SU$ theories, since the ratio of the diagrams becomes $2 \times 2^{4/2} \times 2^{-3}= 1$.

Now we can easily generalize the argument to all orders in perturbation theory, which at large $N_{c}$ means generalizing to all planar gluon tilings of the quark loop. A general diagram has $V_1$ quark-gluon vertices, $V_3$ three-gluon vertices and $V_4$ four-gluon vertices, and $N_g$ gluon propagators. The ratio of the leading order diagrams in $SO$ to $SU$ will have a factor of $2^{\frac{V_1 + V_3 + 2V_4}{2}}$ from the coupling constants. Whilst in $SU(N_{c})$ the orientation of any three or four gluon vertex is fixed by the external quark loop, we can construct a planar diagram from either orientation in $SO(N_{c})$ by using the twisted propagators. The number of equivalent planar diagrams is therefore given by $2^{V_3 + V_4}$. Taking into account the factors from the propagators the overall ratio is therefore: 
\be
 \frac{SO}{SU} = 2^{\frac{4V_4 + 3 V_3 + V_1}{2} }\times 2^{-N_g}.
\ee
But of course each gluon propagator must start and end on a vertex, so
\be
 2 N_g =   4 V_4 + 3 V_3 + V_1 ,
\ee
and we see that the ratio of leading order diagrams is always unity. Thus we have established planar equivalence for common-sector mesons to all orders in perturbation theory, without invoking the idea of orbifold projections. 

\subsection{Baryons at large $N_{c}$}
\label{sec:BaryonEquivalence}

In this section we generalize the large $N_c$ equivalence to baryon correlation functions. Just as in the meson case we expect large $N_{c}$ equivalence to apply to a subsector of the baryons in the $SO(N_c)$ theory. Clearly we can only directly compare correlation functions that are gauge invariant in both theories, and therefore the natural candidate for the common sector of $SO(N_c)$ are the baryons with the quantum numbers of $N_{c}$ quarks (or $N_{c}$ antiquarks). We will first review the diagrammatics of these baryons in $SU(N_c)$ gauge theory, before showing the equivalence with the common sector of the $SO(N_c)$ theory. 

The discussion of baryons in the large $N_{c}$ limit is more complicated than the analysis of mesons. Since a baryon contains $N_{c}$ quarks, the diagrams one can draw in perturbation theory depend on $N_{c}$, and analyses of baryons in terms of Feynman diagrams at large $N_{c}$ are not straightforward.  Nevertheless, Witten showed in his seminal paper\cite{Witten:1979kh} that simple scaling relations do hold for baryons, which we now review.   We will focus our attention on the simplest non-trivial observable involving baryons, which is the baryon two-point function.  To keep things as simple as possible, we will discuss the two-point function of the operator
\be
\label{eq:BaryonOp}
J(x) = \Psi^{1}(x)...\Psi^{N_{c}}(x) =
  \frac{1}{N_{c}!} \epsilon_{i_1...i_{N_{c}}}\Psi^{i_1}(x)...\Psi^{i_{N_{c}}}(x) ,
\ee
where $\Psi$ is a single flavour Dirac field, the subscripts shown refer to color, and the Lorentz indices are suppressed.  If one wants to include the Lorentz structure, then the simplest thing to consider is to take each $\Psi^{i}$ be drawn from the same spinor component of $\Psi$. 

Despite the complications of counting, Witten showed that one can draw interesting conclusions by focusing on connected interactions involving  $m$-quarks for $m \ll N_{c}$. The restriction to $m \ll N_{c}$ would be justified if the fermions were heavy, since then $\lambda(m_{Q}) \ll 1$.  If the quarks are light, considering only diagrams with $m \ll N_{c}$ is not formally justified, but gives great qualitative insight.  Finally, the restriction to the connected interactions is based on Witten's key insight that the baryon mass scales as $N_{c}^{1}$, and as a result the disconnected interactions essentially exponentiate the connected ones.  

\begin{figure}
\includegraphics{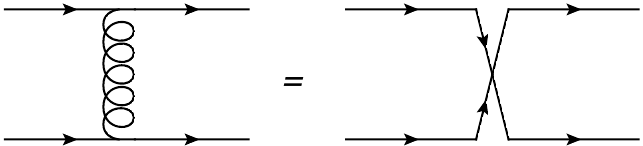}
\caption{\small\emph{Two quarks inside a baryon in $SU(N_{c})$ gauge theory interact at first order by exchanging color. The diagram $\sim N_{c}^{-1}$ but the number of ways of choosing the two quarks from the $N_{c}$ quarks making up the baryon, which scales as $N_{c}^{2}$, means that the overall effect is $\sim N_{c}$.}}
\label{fig:OGESUNBaryon}
\end{figure}

\begin{figure}
\includegraphics{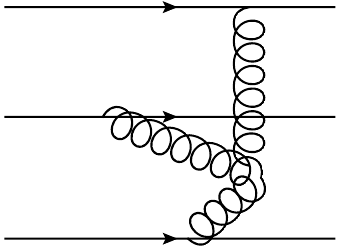}
\caption{\small\emph{A three quark interaction in a baryon that scales as $N_{c}^{-2}$. At large $N_{c}$ there are $N_{c}^3$ quarks that can take part and so the overall effect also $\sim N_{c}$.}}
\label{fig:3GBaryon}
\end{figure}

The first order interaction in perturbation theory corresponds to two quarks in the baryon exchanging a gluon as shown in Fig.~\ref{fig:OGESUNBaryon}. It is simple to work out the color factor. Using the $U(N_{c})$ propagator the only interaction corresponds to the two quarks exchanging color and so we simply get a factor of $1/N_{c}$ introduced by the two vertices. However there are $\frac{1}{2} N_{c}(N_{c}-1)$ pairs of quarks that can exchange the gluon and so the overall amplitude scales as $N_{c}^{1}$. The interesting thing Witten noticed is that the same scaling holds for any leading order interaction with $m \ll N_{c}$. The leading order connected interactions between $m$ quarks scale as  $N_{c}^{1-m}$ - for example Fig.~\ref{fig:3GBaryon} shows a three-quark interaction that scales as $N_{c}^{-2}$. However for $m \ll N_{c}$ there is a combinatorial factor of $\sim N_{c}^{m}$ which counts the number of choices of $m$ quarks to participate in the interaction from the $N_{c}$ available quarks in the baryon. One therefore finds that the contribution of $m$-body diagrams to the baryon two-point function always scales as $N_{c}$.  

As a consequence of these observations, Witten argued that there is a simple large $N_{c}$ limit for baryons which has the baryon mass scaling as $N_{c}$. This scaling suggests that baryons can be viewed as solitons of the meson fields, since $M_{B} = 1/\lambda_{\textrm{meson}}$, where $\lambda_{\textrm{meson}} \sim 1/N_{c}$ is the quartic coupling constant of mesons at large $N_{c}$. Since the common-sector mesons of $SO(N_c)$ are equivalent to the mesons of $SU(N_c)$ gauge theory, we might expect to be able to construct a soliton of common-sector mesons in $SO(N_{c})$ that would be identifiable with the $SU(N_{c})$ baryon. Standard large $N_{c}$ arguments therefore suggest that we should be able to find some equivalence between baryons. Following our approach to mesons, let us investigate this statement by comparing the two-point function of the baryon operator $J(x)$ in the two theories using perturbation theory. 

\begin{figure}
\includegraphics{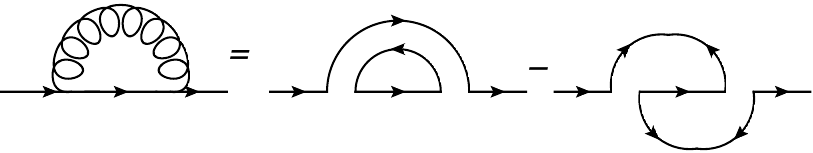}
\caption{\small\emph{At first order the renormalized quark propagator has two contributions in $SO(N_{c})$. The twisted propagator gives a term subleading in $N_{c}$}}
\label{fig:OneBodyBaryon}
\end{figure}

At first order in perturbation theory we have two Feynman diagrams to consider. Firstly a single quark can emit and then reabsorb a gluon - a one-body interaction which is a contribution to the renormalized quark propagator.  There are $N_{c}$ such diagrams, each of which scales as $\mathcal{O} (1)$.  In Fig.~\ref{fig:OneBodyBaryon} we show the color structure of this interaction in the $SO(N_{c})$ theory. The diagram corresponding to the usual propagator agrees precisely with the one in the $SU(N_{c})$ theory, with the factors of $2$ in the coupling constants and propagators canceling. The diagram containing the twisted propagator is suppressed by $N_{c}$ and so we have an agreement at leading order.

\begin{figure}
\includegraphics{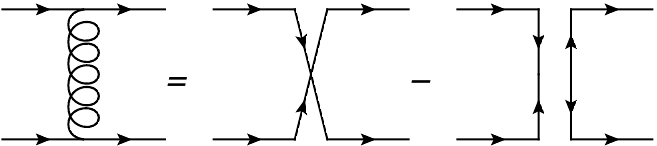}
\caption{\small\emph{In $SO(N_{c})$ there are two ways quarks can interact at first order. However since the quarks inside a baryon always have different colors, this additional term plays no role.}}
\label{fig:OGESONBaryon}
\end{figure}

Second, two quarks of different color can exchange a gluon, shown in Fig.~\ref{fig:OGESONBaryon}. As we noted before in $SU(N_{c})$ this simply corresponds to them swapping color. Once more the first term in the $SO(N_{c})$ propagator reproduces this effect - the factors of 2 working out as before. The additional term in the propagator does not contribute at all - recall it has the color structure $\delta^{ik} \delta_{jl}$. At tree level this term is only relevant for two quarks of the same color, and so at first order it plays no role inside a baryon. 

There are two important observations we can make from this first order equivalence. First recall that the effect of connected baryon interactions scales as $N_{c}$. From our discussion of the renormalized propagator it is clear that there are going to be subleading corrections to these interactions that differ in the two theories. Therefore the statement of large $N_{c}$ equivalence for baryons must apply to \emph{connected} $m$-quark interactions. A second observation is that it is crucial that we are considering interactions between quarks of different colors. For quarks of the same color the effects of the twisted propagator cannot be ignored, and the interactions would not agree in the two theories. 

That this result depends crucially on the baryon structure, i.e. that the quarks have different colors, is non-trivial and suggests we should be able to generalize it to higher orders. To do this we need to show that connected $m$-quark interactions between $m$ different colors will agree at leading order in $N_{c}$. \emph{Given} that the quarks have different colors, the key in showing this result at first order was that the factors of 2 in the propagator and the coupling constants cancel, which are the same factors that were relevant for the meson equivalence. This suggests that the combinatorial factors discussed in the meson sector are also the key to realizing the large $N_{c}$ equivalence of baryons.
 
\begin{figure}
\includegraphics{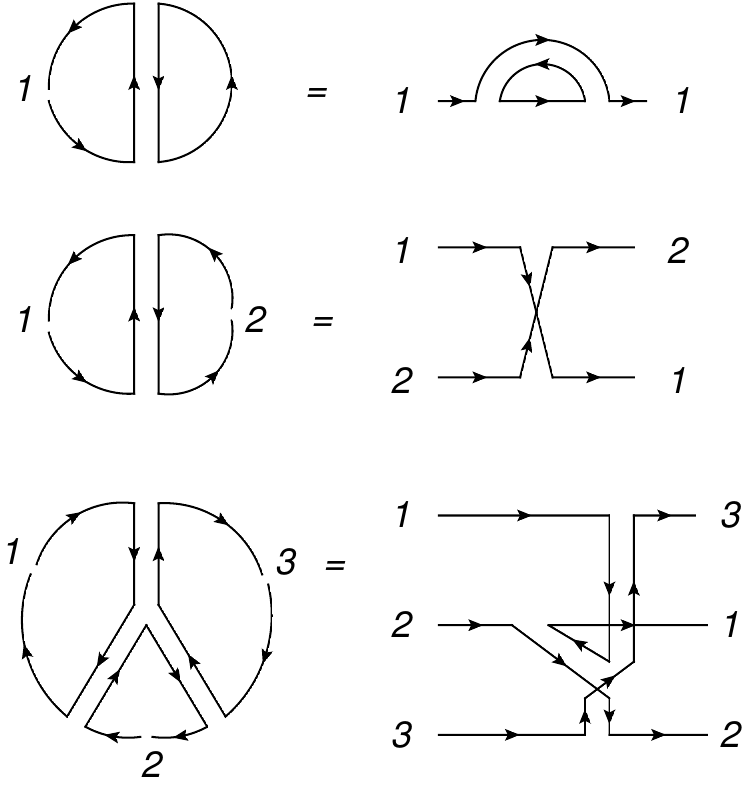}
\caption{\small\emph{Leading order $m$ body interactions for a baryon can be constructed by cutting planar diagrams on $m$ index loops and inserting a different color on each.}}
\label{CuttingPlanarDiagrams}
\end{figure}

That the combinatorics are in fact the same in both cases can be made explicit by using the fact that leading order baryon interactions can be constructed from (common-sector) planar meson diagrams. The procedure, described in (for instance) \cite{Manohar:1998xv}, is as follows. Any leading order connected interaction between $m$ different colored quarks can be constructed by cutting a planar meson diagram on $m$ different color index loops. One then inserts a different color index, for each of the quarks in the interaction, onto each broken loop. The resulting diagram is therefore a consistent color flow for $m$ external quarks of different colors - i.e. a connected baryon interaction. Apart from the $m$ broken loops, the diagram is the same as a planar meson diagram and so scales as $N_{c}^{1-m}$, which is the scaling of a leading order baryon diagram before multiplication by the baryonic combinatorial factor $\sim N_{c}^{m}$.  This process is illustrated for $SU(N_{c})$ diagrams in Fig.~\ref{CuttingPlanarDiagrams}. 

All leading order baryon diagrams can be constructed in this way.  The same argument can be used to construct the leading order baryon diagrams in $SO(N_{c})$. The key point is that since all leading order baryons diagrams can be viewed as being constructed from meson diagrams, the combinatoric factors in each given $m$-order interaction ---  i.e. the numbers of equivalent diagrams, propagators and coupling constants ---  are identical to the meson case. For example, consider the 3-quark interaction of Fig.~\ref{fig:3GBaryon}. As in the meson case illustrated in Fig.~\ref{fig:3GAdditionalSODiagram}, there are four factors of the coupling constant and 3 gluon propagators and so in order to get large $N_{c}$ equivalence we require that there are twice as many leading order diagrams in $SO(N_{c})$. However we know this to be the case because we can construct a baryon diagram from either of two meson diagrams appearing in Fig.~\ref{fig:3GAdditionalSODiagram}.

Thus we see that at the diagrammatic level we have a natural generalization of orbifold equivalence to baryons. In the meson case, we have a leading order equivalence between the common mesons. In the baryon sector the equivalence relates the $SO(N_c)$ baryons with quark number charge $N_{c}$ to the $SU(N_c)$ baryons. We have just seen that the diagrams associated with these common-sector $SO(N_{c})$ baryons match to the equivalent ones for $SU(N_{c})$ baryons, in the sense that the connected diagrams contributing to the baryon two-point function agree at leading order in $N_{c}$. The full two-point function can be interpreted as exponentiating this contribution. In particular we therefore expect the baryon masses in the two theories to agree at leading order. 

In most of this paper we focus on a comparison between the two-point functions of baryons in $SU(N_{c})$ and $SO(N_{c})$ gauge theories.  The fact that agreement between the two theories in the two-point baryon correlation function sector followed simply from the planar equivalence between $SO(N_{c})$ and $SU(N_{c})$ makes it extremely plausible that the same should happen for the higher point functions. At first order in perturbation theory a baryon-baryon interaction will involve one quark from each baryon exchanging a gluon. One must make a choice of one quark from the $N_{c}$ quarks from each baryon to take part in this interaction, and so the number of possible diagrams scales as $N_{c}^2$. The baryon-baryon interaction therefore scales as $\mathcal{O}(N_c)$ just like the baryon mass.  Most ($\sim N_{c}^{2}$) of these possible interactions involve quarks of different colors in the two baryons and therefore will agree between the two theories. Of course, there are also $\sim N_{c}$ diagrams involving interactions between quarks of the same color, and these will not agree between the two theories, but these effects are $1/N_{c}$ suppressed.  Clearly, we can make the same argument for any $m$-quark interaction with $m \ll N_{c}$. Since these are the interactions we can identify as contributing to the leading $\mathcal{O}(N_{c})$ effect, this suggests that the large $N_{c}$ equivalence extends to baryon interactions as well, as is discussed from another perspective in the following section.

\subsection{Baryon equivalence on the hadronic level}
\label{sec:BaryonInteractions}

\begin{figure}
\includegraphics{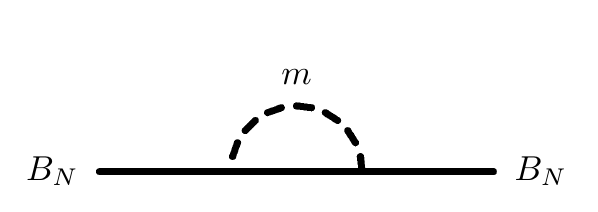}
\caption{\small\emph{Leading-order meson loop contribution to baryon propagator. }}
\label{fig:PropagatorCorrection}
\end{figure}

Standard large $N_{c}$ counting indicates that mesons contribute to the properties of baryons at leading order\cite{Cohen:1992kk}. In particular the baryon two-point function involves leading-order corrections from meson loops, as illustrated in Fig.~\ref{fig:PropagatorCorrection}, whilst baryon-baryon scattering can be understood via the exchange of mesons. Due to the presence of b-mesons in the $SO(N_{c})$ theory one may worry that this raises some doubts about our conclusions based on perturbation theory. In this section we argue that for baryons with quark number $N_{c}$ the analogous processes involving b-mesons are suppressed at large $N_{c}$.

\begin{figure}
\includegraphics{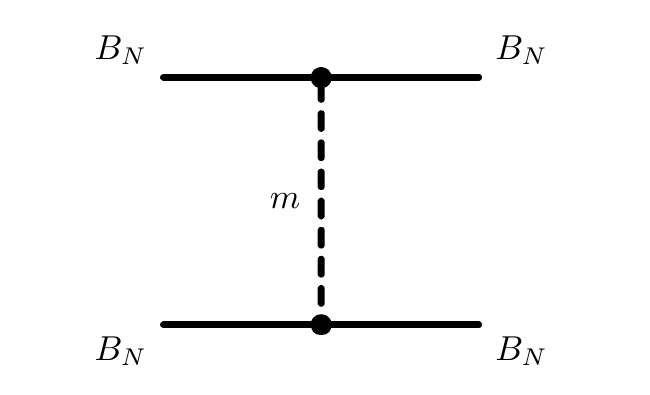}
\caption{\small\emph{One-meson-exchange contribution to baryon-baryon scattering amplitude}}
\label{fig:OneMesonExchange}
\end{figure}

Let us start by briefly reviewing the large $N_{c}$ behavior of the baryon-meson coupling constant $g_{mBB}$\cite{Witten:1979kh} in $SU(N_{c})$ gauge theories. The fastest way to find the large $N_{c}$ scaling of $g_{mBB}$ is to consider the contribution of one meson exchange to the baryon-baryon scattering amplitude which is proportional to $g_{mBB}^{2}$ and is illustrated in Fig.~\ref{fig:OneMesonExchange}. One-meson exchange between baryons can be thought of as an exchange of constituent quarks from one baryon to the other.  It is important to recall that thanks to the Levi-Civita tensor involved in the color structure of baryons, each quark color occurs precisely once in a baryon.  Hence if a quark with color $c$ --- for which there are $N_{c}$ choices --- in one baryon is to be exchanged with one of the $N_{c}$ quarks in another baryon, the new quark must have the same color $c$.  As a result, the one-meson-exchange diagram will scale as $g_{mBB}^{2} \sim N_{c}^{1}$, and as a result we see that $g_{mBB} \sim N_{c}^{1/2}$.  This is in sharp contrast to the large $N_{c}$ scaling of meson-meson coupling constants, which are such that in purely mesonic processes, meson loops are suppressed at large $N_{c}$.  On the other hand, the scaling $g_{mBB} \sim N_{c}^{1/2}$ leads to the naively surprising fact that a class of meson loop diagrams contribute to baryon properties at \emph{leading order} in the large $N_{c}$ limit.

The argument for the large $N_{c}$ scaling of $g_{mBB}$ is exactly the same in an $SO(N_{c})$ gauge theory.  However, the presence of b-mesons $b$ in the $SO(N_{c})$ theory also allows a b-meson-baryon coupling constant.  If it were to be the case that the b-meson-baryon coupling constant scales as $\sim N_{c}^{1/2}$, one would worry that b-meson loops would contribute to baryon properties in $SO$ theories, raising some doubts about the conclusions of the previous section.  

As we discussed previously, baryons in the $SO(N_c)$ theory can be composed of a mixture of quarks and antiquarks - we shall label these operators as $B_q$ where $q$ indicates the net quark number.  Our equivalence predicts that the correlation functions of baryons composed entirely of quarks, the $B_{N_c}$ operators, should match between the $SO(N_{c})$ and $SU(N_{c})$ theories. The interactions we might be concerned about would involve b-meson loops dressing a $B_{N_{c}}$ propagator, or  $B_{N_{c}}$ - $B_{N_{c}}$ interactions involving exchanges of an even number of b-mesons.   Quark number conservation implies that a baryon with quark number charge $N_{c}$ which emits a b-meson with charge $+2$ must turn into a baryon with quark number charge $N_{c}-2$ (i.e. a baryon composed of $N_{c}-1$ quarks and an antiquark).  Hence we denote the relevant coupling constant as $g_{bB_{N_{c}}B_{N_{c}-2}}$.  To understand the importance of these processes, the crucial issue is the large $N_{c}$ scaling of $g_{bB_{N_{c}}B_{N_{c}-2}}$. 

\begin{figure}
\includegraphics{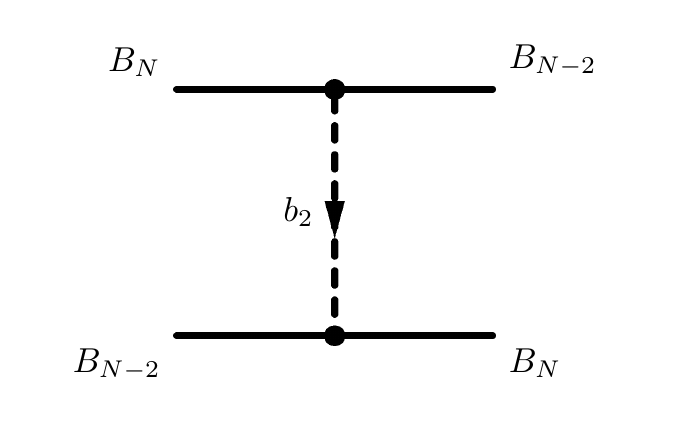}
\caption{\small\emph{One-b-meson-exchange contribution to scattering amplitude of a baryon $B_{N}$ with a $B_{N-2}$ baryon, with the subscript indicating the quark number charge.}}
\label{fig:Oneb-mesonExchange}
\end{figure}

To work out the large $N_{c}$ scaling of $g_{bB_{N_{c}}B_{N_{c}-2}}$, then, one can consider the one-b-meson exchange scattering amplitude of a $B_{N_{c}}$ baryon and a $B_{N_{c}-2}$ baryon, illustrated in Fig.~\ref{fig:Oneb-mesonExchange}. As in the above argument for the scaling of $g_{bBB}$, to find the scaling of $g_{bB_{N_{c}}B_{N_{c}-2}}$ it is sufficient to consider the scaling of the quark-exchange diagram.  If we wish to constuct a diagram that can viewed as the exchange of a b-meson then we must have a net quark number flow. This means that the exchange \emph{must} involve the antiquark in $B_{N_{c}-2}$ and a quark of the same color from $B_{N_{c}}$.  There is therefore no combinatorial factor of $N_{c}$ for the diagram.  Hence $g_{bB_{N_{c}}B_{N_{c}-2}} \sim 1$ at large $N_{c}$, in contrast to $g_{mB_{N_{c}}B_{N_{c}}}$, and b-meson loops make \emph{subleading} contributions to the properties of common-sector baryons.  Futhermore we can also see to that two-b-meson exchange between common-sector baryons scales as $\mathcal{O}(1)$, and so is suppressed relative to meson exchange. These considerations suggest that our large $N_{c}$ equivalence is consistent with the presence of b-mesons in the $SO$ theory. 

\section{Large $N_{c}$ equivalence in $2D$}
\label{sec:2D}

Having discussed how large $N_{c}$ equivalence for baryons works in perturbation theory, we now turn to study $SO(N_{c})$ and $SU(N_{c})$ gauge theories in two dimensions, where we can get some insight into the non-perturbative conditions for large $N_{c}$ equivalence to be valid.  Two-dimensional QCD, often called the 't Hooft model, is dramatically simpler than its higher-dimensional analogues because gauge fields can be made non-dynamical by a judicious choice of gauge fixing conditions.  In his original paper on large $N_{c}$ baryons, Witten used this model to demonstrate the validity of his large $N_{c}$ counting away from the heavy quark limit\cite{Witten:1979kh}. We go through a very similar analysis in order to demonstrate the equivalence between baryons in $SO(N_{c})$ and $SU(N_{c})$ gauge theory.

In Coulomb gauge $A_{1}=0$, one can integrate out the gauge fields in two-dimensional large $N_{c}$ gauge theories with fermions, yielding theories of fermions interacting through a non-local four-fermion interaction (the color-Coulomb interaction).  The structure of the four-fermion interaction terms differs between the two theories thanks to the different color structures of the gluon propagators.  Calling the four-fermion terms in the action $S_{1}, S_{2}$, we write the action for $SO(N_{c})$ gauge theory in the form
\be 
S_{SO} = \int dx dt\, \left[ \bar \Psi D \Psi \right]+ S_{1} + S_{2}
\label{eq:SO2DAction}
\ee
where $D = \slashchar{D} +M$, whilst the corresponding action for $SU(N_{c})$ gauge theory at large $N_{c}$ is
\be
S_{SU} = \int dx dt\, \left[ \bar \Psi D \Psi \right]+ S_{1}.
\label{eq:SU2DAction}
\ee
$S_{1,2}$ are defined as
\begin{align}
S_{1} &= - \frac{\Lambda}{N_{c}} \int dxdydt \, (\bar \Psi_i \gamma^0 \Psi^j) (x,t)(\bar \Psi_j \gamma^0 \Psi^i) (y,t)|x-y| \nonumber \\
S_{2} &= \frac {\Lambda}{N_{c}} \int dxdydt \, (\bar \Psi_i \gamma^0 \Psi^j) (x,t)(\bar \Psi_i \gamma^0 \Psi^j) (y,t)|x-y| \nonumber ,
\end{align}
and $\Lambda$ has mass dimension $-2$ and is proportional to the 't Hooft couplings in the $SU(N_{c})$ and $SO(N_{c})$ theories.   For the $SU(N_{c})$ theory the parameter $\Lambda = \lambda$, whilst for $SO(N_{c})$ it is $2\Lambda = \lambda$.   Large $N_{c}$ equivalence therefore relates the actions $S_{SU}$ and $S_{SO}$ with the same $\Lambda$. The key point in what will follow is that the interaction $S_1$ has the structure $(\bar{\Psi}_{i} \Psi^{i})(\bar{\Psi}_{j} \Psi^{j})$ in color space whereas $S_2$ is of the form $(\Psi^{i} \Psi^{i})(\bar{\Psi}_{j} \bar{\Psi}_{j})$. The extra term in $S_{SO}$ has the color structure of a b-meson-b-meson interaction and as a result we will be able to argue that it does not affect the properties of common-sector mesons at leading order.

The standard trick to deal with four fermion interactions is to linearise them via the introduction of auxiliary fields. Following Witten we use a non-local version of this trick to linearise the action via the introduction of a meson field $m(x,y,t) \sim \bar \Psi_i(x,t) \Psi^i(y,t)$ and a b-meson field $b(x,y,t) \sim \Psi^i(x,t) \Psi^i(y,t)$.  Strictly speaking $m$ and $b$ should be introduced as a matrix of fields to represent the Dirac structure of the above interaction. However, carrying around the Dirac indices would clutter the presentation without adding substantively to it, since the crucial role is played by the color indices.  Thus in what follows we we suppress the Dirac indices for simplicity; they are easy to restore should one wish to do so. 
 
Upon integrating in auxiliary fields, we get an equivalent but different-looking action to Eq.~\eqref{eq:SO2DAction}, which can be (schematically, since the Lorentz structure is being suppressed) written as
\be S_{SO} = \int dxdt \bar \Psi D \Psi + N_{c}S_{\textrm{free}} + S_{1} + S_{2} 
\ee
where
\begin{align}
S_{\textrm{free}} &= \int dxdydt\,  m(x,y,t) m^{*}(x,y,t) + b(x,y,t) b^{*}(x,y,t)  \,,   \nonumber \\
S_1 = &\sqrt{\Lambda} \int dxdydt\, m(x,y,t) \bar \Psi_i(x,t)\Psi^i(y,t) \sqrt{|x-y|} + h.c. ,\nonumber \\
S_2 = &\sqrt{\Lambda} \int dxdydt\, b^{*}(x,y,t) \Psi^i(x,t)\Psi^i(y,t) \sqrt{|x-y|} + h.c.  \nonumber
\end{align}
The relevant action for $2D$ $SU(N_{c})$ QCD is the same but with the b-meson fields set to zero. 

\subsection{2D Meson equivalence}

Before moving on to baryons it is instructive to see how the familiar results of large $N_{c}$ equivalence are reproduced non-perturbatively in the 't Hooft model. Let us choose the simplest operator that will display orbifold equivalence - the renormalized quark propagator $\langle \bar \Psi(x) \Psi(0) \rangle$.  This can be calculated in the $SO(N_{c})$ gauge theory via the path integral 
 \begin{align}
\langle \bar \Psi(x) \Psi(0) \rangle &=  Z^{-1} \int d[m]d[b] d[\Psi] d[\bar{\Psi}] \left( \right. \nonumber\\
&\left. \bar \Psi(x)\Psi(0) e^{iS_{SO}} \right) 
\end{align}
After integrating over the fermions, $\bar \psi(x)\psi(0)$ becomes replaced by the quark propagator in the background field $D^{-1}(x, m, b)$. We also pick up a factor of the partition function of the fermions in the background fields $m$ and $b$. In the $SU(N_{c})$ case with $b=0$, this corresponds (schematically) to $\det(\gamma^{\mu}p_{\mu} + M - \sqrt{\Lambda} m)$.  However, since the $b$ field couples to $\psi^{T} \psi$, the result of the fermion integration in the $SO$ theory, with $b \neq 0$, is the Pfaffian $\Pf(C K)$ instead of the determinant $\det(D)$, with
\begin{align}
K = 
\left(
\begin{array}{ccc}
  \gamma^{\mu}p_{\mu} + M - \sqrt{\Lambda} m &  \sqrt{\Lambda} b          \\
 \sqrt{\Lambda} b^{\dag} &   \gamma^{\mu}p_{\mu}  + M - \sqrt{\Lambda}m       
\end{array}
\right).
\end{align}
and $C$ the charge conjugation matrix. Up to a sign, $\Pf(C K) = (\det K)^{1/2}$, and we can write $\Pf(C K) = \exp(\frac{1}{2} \Tr \log K)$. The fact that the original action is diagonal in color space means that the color trace simply gives $N_{c}$.  Thus we are left with 
\be 
\langle \bar \Psi(x)\Psi(0) \rangle = \int d[m] d[b] D^{-1}(x,m,b) e^{iS_{\textrm{eff}}(m, b)},
\ee
where
\begin{align}
\label{eq:MesonEffectiveAction}
S_{\textrm{eff}}^{M}(m, b) &= N_{c} \left[ \frac{1}{2} \Tr \log K(m, b) + S_{\textrm{free}}(m, b) \right]
\end{align}
is the effective action on the meson fields. The key point is that having integrated out the fermions the only dependence on $N_{c}$ is the factor in front of effective action which we have explicitly shown. This means that it is now easy to find the large $N_{c}$ limit of the theory:  for $ N_{c} \gg 1 $ the integral can be evaluated by a saddle-point/stationary-phase approximation.  There will exist some set of field configurations $(m^0, b^0)$ that extremize $S^M_{\textrm{eff}}$, corresponding to (possibly unstable, for some cases) phases of the theory.  Given one particular extremizing configuration, to the leading order in the $1/N_{c}$ expansion,  the quark propagator in the associated phase can simply be evaluated in this background field:
\be 
\langle \bar \Psi(x)\Psi(0) \rangle_{SO} = D^{-1}(x, m^0, b^0) .
\ee

In order to have a large $N_{c}$ equivalence it is clear that we need the $SO$ theory to have a \emph{stable} saddle point of the form $(m^0, b^0) = (m^{qcd}, 0)$ where $m^{qcd}$ is  the stable saddle points in the $SU(N_{c})$ theory\footnote{To keep things simple, this paragraph is written assuming that there is only one stable saddle point in e.g. the $SU$ theory, but this need not be true.  It is not hard to adjust the discussion to take into account the possibility that there are metastable phases in the two theories.}. Since the b-meson field is charged under $U(1)_Q$ and the effective action on this subspace is the same as in $SU(N_{c})$, we are guaranteed that  the above configuration will be an extremum point of $S_{\textrm{eff}}$, but we are not a priori guaranteed that it is a stable extremum.   \emph{If} the saddle point is stable the effective actions for the two theories coincide and therefore so do the quark propagators to leading order in $N_{c}$. We therefore have a large $N_{c}$ equivalence, provided that the relevant saddle point in the $SO(N_{c})$ theory is stable. However, the value of b-meson field at the saddle point represents its vacuum expectation value (to leading order in $N_{c}$). Consequently the question of the stability of this saddle point is equivalent to whether the $U(1)_{Q}$ symmetry is `spontaneously broken'\footnote{Spontaneous symmetry breaking is strictly speaking impossible in 2D theories at finite $N_{c}$, but there is a sense in which symmetry breaking becomes possible as $N_{c}\to \infty$ in 2D theories, as explained in e.g.~\cite{Witten:1978qu}. }. In the $U(1)_{Q}$ unbroken phase, the b-mesons cannot pick up a vacuum expectation value by definition, and so the additional interaction in the $SO$ theory does not renormalise the propagator at leading order in $N_{c}$.  For zero chemical potential, we expect that the symmetry will be preserved thanks to the Vafa-Witten theorem\cite{Vafa:1983tf}, which forbids spontaneous breaking of vector-like symmetries such as $U(1)_{Q}$ in the class of theories we consider here. Hence we expect that large $N_{c}$ equivalence will hold at zero chemical potential.  If a $U(1)_{Q}$ chemical potential is turned on, a detailed analysis is necessary to determine the realization of the symmetry, and if necessary one can introduce deformations to protect $U(1)_{Q}$\cite{Cherman:2010jj,Cherman:2011mh}. 

The effective action in \eqref{eq:MesonEffectiveAction} allows one to explore the correlation functions of the $m$ and $b$ fields, since at large $N$ their dynamics are governed by fluctuations about the saddle point. If we absorb a factor of $\sqrt{N_{c}}$ into the fields we can see that the $p$-field coupling constant in $S_{\textrm{eff}}$ scales as $N_{c}^{1-p/2}$. This is simply an explicit realization of the well known idea that confining large $N_{c}$ gauge theories become weakly interacting theories of meson/b-meson fields. The above scaling ensures that to leading order in $N_{c}$ we can evaluate the meson correlation functions using the tree level approximation to $S_{\textrm{eff}}$. This tree-level argument was noted by Cherman and Tiburzi \cite{Cherman:2011mh} as a heuristic explanation of how large $N_{c}$ equivalence works at the hadronic level. For example, if one considers an $m + m \longrightarrow m + m$ scattering amplitude, the leading order terms correspond to a contact interaction and tree-level intermediate meson exchanges due to the $3$-meson coupling. So long as $U(1)_Q$ is conserved,  charge conservation implies only common-sector mesons, which have zero $U(1)_{Q}$ charge, can appear on internal legs in tree-level diagrams, and so b-mesons make no contribution to these correlators at leading order. As shown by our construction of the above effective action, this is precisely how orbifold equivalence is realised in the 't Hooft model. As we remarked above the meson coupling constants are the same in  the $SO(N_{c})$ and $SU(N_{c})$ effective actions, and so orbifold equivalence is realized for these correlators so long as $U(1)_{Q}$ is unbroken.

\subsection{2D Baryon equivalence}

The path integral treatment we used in the last section can be repeated to analyse baryons. We now wish to evaluate the baryon two-point function: 
\begin{align}
 \langle J^{\dagger}(x) J(0) \rangle =  Z^{-1} \int d[m] d[b] \, d[\Psi] d[\bar{\Psi}] \,J^{\dagger}(x) J(0) e^{iS_{SO}} ,
\end{align}
where $J$ is given in Eq.~\eqref{eq:BaryonOp}. Once more we integrate out the fermions.  Since the action is diagonal in the color indices $J^{\dagger}(x)J(0) = \psi^{\dagger}_{N_{c}}(x) \ldots \psi^{\dagger}_1(x) \psi^1(0) \ldots \psi^{N_{c}}(0)$ becomes replaced by $\left(D^{-1}(x, m, b)\right)^{N_{c}}$, yielding 
\be
 \langle J^{\dagger}(x)J(0) \rangle = Z^{-1} \int d[m] d[b] (D^{-1}(x, m, b))^{N_{c}} e^{iS^M_{\textrm{eff}}(m,b)}  \nonumber
 \ee
The saddle-points of $S^M_{\textrm{eff}}$ do not directly determine the baryon correlation functions because of the explicit $N_{c}$ dependence in the baryon operator $J$.  However, following Witten \cite{Witten:1979kh}, we can define a new effective action in the baryon sector:
\begin{align}
\langle J^{\dagger}(x)J(0) \rangle = \int d[m] d[b] e^{iN_{c}S^B_{\textrm{eff}}(m, b, x)} \nonumber
 \end{align}
where
\begin{align}
S_{\textrm{eff}}^{B}(m, b, x) &= \left. \bigg[ \log (D^{-1}(x, m, b)) +  \right. \nonumber\\
&\left. \frac{1}{2} \Tr \log \tilde{M}(m, b) + S_{\textrm{free}}(m, b) \right. \bigg] ,
\end{align}

For $N_{c} \gg 1$ we can evaluate the baryon two-point function as the saddle point approximation to $S^B_{\textrm{eff}}$. Ignoring numerical factors, we find 
\be
 \langle J^{\dagger}(x)J(0) \rangle =Z^{-1} \frac{e^{iN_{c}S^B_{\textrm{eff}}(m^{B}, b^B, x)}}{\sqrt{\det(d^2 S^{B}_{\textrm{eff}})}} 
 \ee
where $(m^B, b^B)$ are the coordinates of the saddle point and $d^2 S^{B}_{\textrm{eff}}$ is the matrix of second derivates of the effective action at the saddle point. This term is necessary because it will not be cancelled by anything in the partition function. As Witten noticed this analysis reproduces the large $N_{c}$ counting we saw in perturbation theory. There is an order $N_{c}$ contribution to the baryon mass coming from the saddle point, and then order $N_{c}^0$ corrections corresponding to meson fluctuations about it. 

Since $S_{\textrm{eff}}^B$ still possesses a $U(1)_Q$ symmetry there is once again a saddle-point configuration with $b^B = 0$ at which the effective action is the same as in $SU(N_{c})$ gauge theory.  Expanding about this saddle-point corresponds to the baryon equivalence we saw in perturbation theory. Assuming the saddle point is stable, the agreement of the saddle-point action implies the baryon masses agree at leading order, but there will be discrepancies at order $N_{c}^0$. 

The question of the stability of the $b^{B}=0$ saddle point is more subtle in the baryon case than in the meson case.   To see whether the $U(1)_{Q}$ symmetry is preserved at the baryon saddle-point, one would have to study the meson and b-meson fluctuations around the baryon solution, and we leave a study of this to future work.  It is conceivable that the $b^{B}=0$ saddle point may be unstable even if the $b^{M}=0$ saddle point is stable, since a baryon background field affects the meson and b-meson spectrum.  If that turns out to be the case, however, it is plausible that one could construct a deformed version of the the $SO(N_{c})$ theory to which the equivalence would apply by adding deformation terms to the action which protect the $U(1)_{Q}$ symmetry and prevent b-meson condensation as was discussed in the 4D context in \cite{Cherman:2010jj,Cherman:2011mh}.  

\section{Conclusions}
\label{sec:Conclusions}

In this paper we have shown that baryons can be naturally incorporated into the framework of large $N_c$ equivalence provided one works with gauge theories of the same number of colors.  In the meson sector large $N_c$ equivalence takes the form of a leading order equivalence between the correlation functions of the mesons common to both $SO(N_{c})$ and $SU(N_{c})$.  We showed that for common-sector correlation functions, the two theories match to all orders in perturbation theory, which was known previously.  The novel step in this paper is the generalization of this argument to the baryon sector, where we showed that the diagrams determining e.g. the masses of baryons agree to all orders in perturbation theory to leading order in the $1/N_{c}$ expansion. For light quarks, the perturbative analysis is only meant to be suggestive, but for very heavy quarks the theory becomes weakly coupled since the relevant coupling becomes $\lambda(m_{Q}) \ll 1$, and hence perturbation theory becomes a reliable approximation.  The all-orders agreement of the perturbative expansions of the two theories immediately implies that the recent calculation of the baryon mass in the large $N_{c}$ and heavy quark limits for $SU(N_{c})$ QCD described in \cite{Cohen:2011cw} also applies to the $SO(N_{c})$ theory.

To get some insight into the extent to which the implications of the perturbative arguments apply for light quarks, we analysed the non-perturbative conditions necessary for the equivalences to hold using the 't Hooft model. In two dimensions, the meson sector orbifold equivalence was shown to hold provided the quark number symmetry $U(1)_Q$ is unbroken at large $N_{c}$.  The realization of the $U(1)_{Q}$ symmetry also turned out to be the critical issue for whether the equivalence holds in the baryon sector.

There are many directions for future work.  While we chose not to use it in our analysis here, it would be nice to find some way to adapt the machinery of orbifold projections to the study of baryon-sector observables. This may allow one to understand the conditions for baryons to be in the common sector of large $N_{c}$ equivalences in general,  rather than checking it in particular cases, as we did here in the context of $SO(N_{c})/SU(N_{c})$ equivalence.  Another direction for future work is to do more detailed studies of the 't Hooft model, since $2D$ QCD-like theories provide a uniquely tractable case where large $N_{c}$ equivalences can be explored non-perturbatively.    Our exploratory study already revealed that, reassuringly, the symmetry-realization conditions identified in \cite{Kovtun:2003hr,Kovtun:2004bz} as the necessary and sufficient conditions for large $N_{c}$ equivalence for theories with matter in two-index representations play the same role in the theories discussed here, with matter in the fundamental representation.  It is important to develop techniques to check whether this continues to be the case in higher dimension.  In such cases analytic methods are not available for many observables of interest, but given the encouraging results we have obtained thus far, it would also be exciting to begin comparing the physics of $SU(N_{c})$ and $SO(N_{c})$ theories using lattice Monte Carlo methods. 

\acknowledgements{
We thank Tom Cohen and Masanori Hanada for discussions, and the anonymous referee for feedback that improved the presentation. MAB thanks Matt Wingate for advice and encouragement and Kenny Wong for comments on an early version of the manuscript. MAB is supported by an STFC studentship. 
}

\bibliographystyle{apsrev4-1}
\bibliography{orbifoldingNoArxiv, orbifoldingEFT} 

\end{document}